\documentclass[aic]{iosart2x}
\pdfoutput=1 

\pubyear{2022}
\volume{0}
\firstpage{1}
\lastpage{1}
\begin{document}
\begin{frontmatter}

\title{From Intelligent Agents to Trustworthy Human-Centred Multiagent Systems}
\runtitle{From Intelligent Agents to Trustworthy Human-Centred Multiagent Systems}

\begin{aug}
\author{\inits{M.}\fnms{Mohammad} \snm{Divband Soorati}\ead[label=e1]{M.Soorati@soton.ac.uk}}
\author{\inits{E.H.}\fnms{Enrico H.} \snm{Gerding}\ead[label=e2]{eg@ecs.soton.ac.uk}}
\author{\inits{E.}\fnms{Enrico} \snm{Marchioni}\ead[label=e3]{E.Marchioni@soton.ac.uk}}
\author{\inits{P.}\fnms{Pavel} \snm{Naumov}\ead[label=e4]{P.Naumov@soton.ac.uk}}
\author{\inits{T.J.}\fnms{Timothy J.} \snm{Norman}\ead[label=e5]{T.J.Norman@soton.ac.uk}}
\author{\inits{S.}\fnms{Sarvapali D. } \snm{Ramchurn}\ead[label=e6]{sdr@ecs.soton.ac.uk}}
\author{\inits{B.}\fnms{Bahar} \snm{Rastegari}\ead[label=e7]{B.Rastegari@soton.ac.uk}}
\author{\inits{A.}\fnms{Adam} \snm{Sobey}\ead[label=e8]{ajs502@soton.ac.uk}}
\author{\inits{S.}\fnms{Sebastian} \snm{Stein}\ead[label=e9]{ss2@ecs.soton.ac.uk}}
\author{\inits{D.}\fnms{Danesh} \snm{Tarpore}\ead[label=e10]{D.S.Tarapore@soton.ac.uk}}
\author{\inits{V.}\fnms{Vahid} \snm{Yazdanpanah}\ead[label=e11]{v.yazdanpanah@soton.ac.uk}%
\thanks{Corresponding author: \printead{e11}}}
\author{\inits{J.}\fnms{Jie} \snm{Zhang}\ead[label=e12]{Jie.Zhang@soton.ac.uk}}

\address{Agents, Interaction and Complexity Research Group, School of Electronics and Computer Science,  \orgname{University of Southampton},
\cny{UK}
}
\end{aug}

\begin{abstract}
The Agents, Interaction and Complexity research group at the University of Southampton has a long track record of research in multiagent systems (MAS). We have made substantial scientific contributions across learning in MAS, game-theoretic techniques for coordinating agent systems, and formal methods for representation and reasoning. We highlight key results achieved by the group and elaborate on recent work and open research challenges in developing trustworthy autonomous systems and deploying human-centred AI systems that aim to support societal good. 
\end{abstract}


\end{frontmatter}

\section{Introduction}

Multiagent systems (MAS) is a dominant trend in artificial intelligence, distributed computing, and computational economics that is having broad impact across society and the economy. The Agents, Interaction, and Complexity Research Group (AIC) in the School of Electronics and Computer Science at the University of Southampton is a leading team contributing both to fundamental theory and practical application. In this whistle-stop tour of our research we focus on four key clusters. The first concerns game-theoretic mechanisms in multiagent systems (Section~\ref{sec:2}) where we discuss how our research contributes to methods for incentivising good collective decisions, fair allocation of resources, auctions and negotiation. All these techniques are centred around how we can make optimal, fair, or (at least) good decisions in systems where disparate parties have conflicting preferences and values. Our second theme explores our contributions to formal representation and reasoning methods, with emphasis on reasoning about multiagent systems and reasoning under uncertainty (Section~\ref{sec:3}). We have particular strengths in probabilistic and possibilistic reasoning, representing and reasoning about norms of behaviour in multiagent systems, responsibility reasoning both in terms of how responsibilities change as agents act and interact and how we can rigorously assess responsibilities for past behaviour, and argumentation-based reasoning. Our third cluster of research is trustworthy autonomous and multiagent systems (Section~\ref{sec:4}), which is research that has centred around major investments of research funding since 2005 including the current UKRI Trustworthy Autonomous Systems Hub. We have investigated the end-to-end design, development and evaluation of systems for energy efficiency applying game-theoretic mechanisms for saving energy in smart homes and flattening load on the energy grid. Research on robust and resilient swarm robotics and human-swarm interaction contributes to understanding how to develop and deploy trustworthy MAS in complex dynamic environments. Safety is also a key element in this trustworthy systems and we have developed mechanisms for safe multi-agent reinforcement learning, lifetime policy reuse and active learning with human involvement. Our final theme brings together all the key elements of our research vision as a group around human-centred AI and multiagent systems (Section~\ref{sec:5}), taking a multidisciplinary approach to key research challenges. We have made major contributions to MAS research on meaningful consent and privacy, and the design of progressive systems for social good. Applications in urban mobility and smart cities including human-centred internet of vehicles and human-centred smart grid are important highlights here.

\section{Game-Theoretic  Mechanisms in MAS}\label{sec:2}

In this section, we review our line of research on game-theoretical mechanisms in multiagent systems. This includes matching mechanisms, approaches rooted in auction theory, and negotiation-based techniques. We conclude the section with open challenges and future directions in this area. 

\subsection{AIC's Research Agenda on Game-Theoretical Mechanisms in MAS}
\label{sec:mechanisms} 
A major challenge in multiagent systems is developing appropriate mechanisms for making collective decisions, e.g. through voting, and allocation of scarce resources, e.g. through auctions, negotiation and matching. These mechanisms take into account agent preferences and typically aim to have certain desirable properties. These include system-level objectives such as achieving globally  optimal solutions based on the agents' preferences, as well as considering individual agents by ensuring  fair outcomes. Typically, mechanisms incentivise truthful reporting of preferences, such that there is no incentive for agents to manipulate the mechanism by misreporting their preferences to the decision maker. Finally, for mechanisms to be practical and to scale to realistic settings, they need to be computationally tractable. 

In the AIC group we investigate both theoretical solutions, as well as practical applications of the mechanisms, often in collaboration with industrial partners. Applications that have been investigated include kidney exchange markets, shipping routing and optimisation, electric vehicle charging, ride sharing, and privacy management. In the remainder of this section, we provide an overview of the matching, auction and negotiation mechanisms developed, their properties, and their practical application.

\paragraph{Matching Mechanisms} Matching under preferences includes problems where agents have preferences over other agents, or objects, and the goal is to match them together while ensuring some desirable properties. Such problems occur widely in practice with examples ranging from school-choice and kidney exchange markets to allocation of computational resources. In most cases, our ideal is to match as many agents as possible. However, in order for the matching to be sustainable and acceptable to the agents/society, one or more other properties need to be satisfied, depending on the application. Stability, Pareto optimality and envy-freeness are arguably the most desirable properties. As in other mechanism design settings, truthful reporting is also desired. The AIC group has contributed to the area of matching under preferences in several directions. The state of the art has been extended to more complex and realistic settings; e.g. by modelling uncertainty in agents' preferences~\cite{azizetal19,azizetal20,azizetal22}. The trade-off between the size of the matching and ensuring truthfulness in settings with one-sided preferences has been studied in \cite{krystaetal,cheklarovaetal}. To circumvent NP-hardness results and inspired by realistic scenarios, correlation between agents' preferences has been exploited in \cite{MeeksRastegari20} leading to tractable solutions in parameterised settings. 

\paragraph{Auction-Based Approaches} 
Auction-like mechanisms are commonly used in agent-based applications as a way to allocate scarce resources when agents are self-interested; i.e. when their aim is  to maximise their own utility. In auctions, agents place bids, and the mechanism determines both the allocation of resources to agents, as well as associated payments. Related to this, in the field of mechanism design, the allocation and payment rules are designed in such a way that the agents are incentivised to report their true valuation for resources, which is typically defined as their maximum willingness to pay. Such mechanisms are truthful or dominant-strategy incentive compatible (DSIC). The most famous DSIC mechanism is the Vickrey or second-price auction, where a single item is allocated to the agent with the highest bid, and the agent pays the bid of the highest loser. The extension to combinatorial settings is the famous Vickrey-Clarke-Groves mechanism. 

In the AIC group, auctions and mechanism design have been applied to a number of applications. In particular, we have extended the work by Parkes et al.~\cite{DBLP:conf/sigecom/FriedmanP03,DBLP:conf/sigecom/Hajiaghayi05} around online mechanism design to the domain of energy allocation. In online mechanism design, the resources arrive and need to be allocated over time (as opposed to the one-off allocation of traditional auctions). It is called \emph{online} since the availability of future resources (and perhaps demand as well) is typically unknown or uncertain. Therefore, it combines aspects of (online) scheduling and mechanism design. Energy allocation is a natural application of online mechanism design, since demand and supply is often uncertain (especially with renewable energy), and needs to be allocated immediately (storage is also a form of allocation). In \cite{robu2013online}, the work was first applied to electric vehicle (EV) charging, where EVs come and go, and therefore future demand is unknown. A model-based setting with probabilistic information about future demand was studied in \cite{stein2012model}. The work was later extended to account for renewable energy and variable supply \cite{hayakawa2015online}. Whilst these works considered EVs as energy consumers, mechanisms for vehicle-to-grid, where EVs can also sell back electricity to the grid were considered in \cite{gerding2016online}. This research considers individual vehicle agents, whereas in \cite{perez2018coordination} we consider the coordination between so-called EV aggregators; i.e. agents which control multiple the charging (and discharging) of multiple vehicles, such as in car parks. Other related works use mechanisms to allocate EVs to charging stations~\cite{rigas2020mechanism,gerding2013two,rigas2022} and model queues at charging stations to optimise EV routing~\cite{deweerdt2013,deweerdt2016}. In addition to EV charging, we have developed novel mechanisms for several other domains including grid computing, where tasks are allocated to compute resources~\cite{gerding2009mechanism, stein2009flexible} and, more recently, the related fog/edge computing~\cite{bi2019truthful}, where computing is done close to where data is generated and services consumed. Other application areas include sponsored search~\cite{ceppi2011mechanism,stavrogiannis2014auction} and ride sharing~\cite{zhao2014incentives,DBLP:conf/ijcai/Iwase0G21}. 

\paragraph{Negotiating  Agents} Negotiation is a way for agents to resolve many different types of conflicts of interest \cite{baarslag2017computers}. Whereas auctions are typically used to allocate scarce resources in MAS, negotiation can be used for other types of decision-making tasks. In particular, negotiation typically involves multiple issues or attributes. For example, in a buyer-seller setting, the negotiation can be about price, but also other issues such as warranty and quality of service. Negotiation is also studied in the energy exchange domain, in which case negotiation issues can be the time of day in which energy is exchanged \cite{alam2015scalable}. Another application area studied within the AIC group is privacy permission management \cite{baarslag2016negotiation, baarslag2017automated}. In this work, a buyer of a service negotiates the terms and conditions with the service provider, and this negotiation is assisted by (but not completely replaced with) a negotiation agent. 

Traditionally, the challenge within agent-based negotiation is developing an effective negotiation strategy. As part of this strategy, a typical approach is to infer the opponent's strategy and/or their utility function, so-called opponent modelling. This way an agent can try to exploit the opponent's strategy, but it also enables finding Pareto efficient agreements which are more likely to suit all parties involved. In the AIC group, the focus has been on a different aspect of negotiation, namely preference elicitation. Whereas traditionally it is assumed agents know the preferences of the users they represent, in practice these need to be elicited. In AIC, we have studied approaches where preferences are elicited \emph{during} the negotiation, and there is a tension between minimising user bother and obtaining the best possible deal \cite{baarslag2015optimalDURING}.

\subsection{Open Challenges and Ways Forward} Our ultimate goal is to employ our results in practical settings. Challenges we face in doing so include extending the current results to more realistic scenarios and convincing policy makers and industry to use these mechanisms. Developing close relationships with industry and policy makers is the route to overcome these challenges.

In learning mechanisms then we need mechanisms that can learn over long periods. Current multiagent systems require a set of prior conditions by the user. These approaches predefine the partition of state space and cost/reward, the numbers of agents, goals and prior beliefs to simplify the problem, allowing training and rewards to be set. However, an increasing focus is on dynamic systems, where these characteristics might change on a regular basis or be unknown at the start. These approaches must be memory efficient, with much of the focus on high-compute approaches. They must also be capable of transferring learning between a range of different tasks without catastrophic forgetting. This needs to be in an environment where the agents can be trusted, which becomes more difficult as the operating life of the system increases. 

Within the context of matching, our current solutions are not likely to work in highly dynamic settings where scarce resources have to be quickly matched to agents in need. An example of such a setting is resource allocation when a disaster strikes, in which physical location and limited mobility of emergency resources are critical to allocation decisions. At the same time, routing decisions also need to be made which in turn affect the feasibility and efficiency of the allocations. Therefore, routing and allocation problems need to be solved simultaneously and repeatedly in such highly dynamic settings.

In auctions and mechanism design, the requirement of incentive compatibility is often too restrictive, and typically not  needed in practice. The notion of epsilon incentive compatibility weakens the strict requirements to situations where misreporting the true preferences  lead to limited improvements in terms of individual utility. There are still many open problems of designing such mechanisms with theoretical guarantees for applications such as ride sharing.  

Another area of ongoing work related to resource allocation mechanisms in general is capturing the user preferences (e.g. in the form of a utility function). Typically, it is assumed the agent knows the user's preferences but, in practice, this needs to be elicited, which takes effort. A question is how accurate these preferences need to be for specific applications, how they can be learned efficiently (e.g. by learning from other users and using similarity between users) and what the trade off is between efficiency of the allocation and the accuracy of the preference profile.

\section{Formal Representation and Reasoning in MAS}\label{sec:3}

In this section, we summarise AIC's research agenda on formal and logic-based methods for reasoning about multiagent systems which is mainly focused around probability-based techniques, normative approaches, and on techniques for argumentation-based reasoning and for reasoning about  different forms of responsibility. 

\subsection{AIC’s Research Agenda on Formal Reasoning in MAS}

\paragraph{Probabilistic and Possibilistic Reasoning} Imagine an agent that comes to a fork in a road. There is a sign that says that one of the two roads leads to prosperity and the other to death. The agent must take one of the roads, but does not know which road leads where. Does the agent have a strategy to get to prosperity? On one hand, since one of the roads leads to prosperity, such a strategy clearly exists. Furthermore, the agent  knows that such a strategy exists. On the other hand, the agent does not know what the strategy is and, therefore, does not know how to use the strategy. If a strategy exists, the agent knows that the strategy exists and knows what the strategy is, then we say that the agent has a {\em know-how} strategy. In the past several years, we have studied know-how strategies to maintain~\cite{nt17aamas} and to achieve~\cite{nt18ai,dn18aaai}, with perfect~\cite{nt18aaai} and bounded~\cite{dn20ai} recall, with knowledge about the opponent~\cite{ny21jair}, and in the presence of the information walls~\cite{nz22aaai}. We have also investigated second-order know-how strategies~\cite{nt18aamas} and know-how strategies with known cost of execution~\cite{cn20ai}.

Probability theory is the most common model in the formalisation and treatment of approximate reasoning in computer science and artificial intelligence, and offers the formal basis for a theory of decision and strategic interaction under uncertainty. Game theory is concerned with studying social interactions, modelling strategic situations in which an individual makes choices depending on the choices of others.
In its classical version, the formal treatment of strategic decisions in game theory is based on the probabilistic model of uncertainty.
However, many situations in which inference and decision are required cannot be properly formalised within this framework. Possibility theory can provide a way to model incomplete knowledge that is alternative to probability, can more adequately capture some forms of uncertainty, and has the advantage of accommodating both qualitative and quantitative representations. One of the main goals in the study of multiagent systems in artificial intelligence is the understanding
of the computational behaviour of systems containing self-interested agents making strategic decisions under uncertainty. Possibility theory can offer a rigorous alternative approach to this problem.
One of our research goals is to bring the key concepts of possibility theory to bear in refining the foundations of non-cooperative game theory. A first step in that direction was achieved in \cite{HoMa19} where a foundational study of the theory of possibilistic games was presented. In this work, we investigate possibilistic games with both a qualitative and quantitative approach offering two different notions of equilibrium.
In \cite{HoMa19} we give a full characterisation of the existence of these equilibria, and analysis of their computational properties and show that this approach can
provide new insights on how agents can efficiently select among multiple equilibria in coordination games.

\paragraph{Norms and Normative Reasoning} Norms are constraints on the behaviour of agents in a social context \cite{LuckNormative2013}. In contrast to causal or resource constraints, agents may act in a manner that violates a norm. As with their analogue in legal systems, sanctions may be imposed that depend on the severity of the violation or there may be reparative actions expected, so called contrary-to-duty obligations. Norms are useful in systems where there are explicit rules that (human or artificial) actors are expected to comply with for operational, safety or security reasons. Normative reasoning enables autonomous agents to assess what norms are in force, the impact they may have on (collective) behaviour, and identify and resolve conflicts among norms \cite{vasconcelos2009} or between norms and intended action. A driver driving on the wrong side of the road in order pass a cyclist can be used to illustrate some relevant concepts. Drivers are forbidden to drive on the wrong side of the road, but if they do they are obligated to ensure that there is no on-coming traffic (a contrary-to-duty obligation). Drivers are obligated to pass cyclists with a safe distance. Considering both of these norms and the goal to drive at reasonable speeds, normative reasoning of this kind can enable agent to make common-sense decisions that are defensible against some set of behavioural expectations.

\paragraph{Responsibility Reasoning}  Study of responsibility is an interdisciplinary topic that originated in ethics and law and is posed to become a very active research area in Artificial Intelligence due to fast increasing number of decisions that humans delegate to autonomous AI agents~\cite{bluesky,dastani2022responsibility}. There have been two different forms of responsibility for actions proposed in philosophical literature: counterfactual responsibility and seeing-to-it responsibility. An agent is {\em counterfactually} responsible for an event if it took place and the agent had a strategy to prevent it. An agent {\em sees} to an event if the action taken by the agent guaranteed that the event would happen. To bring these definitions of responsibility from the level of philosophical discussions to practice in autonomous system design, they require mathematical rigour and need to capture a variety of important settings~\cite{yazdanpanah2021different}. In AIC we formalised these definitions for strategic games with perfect~\cite{nt19aaai} and imperfect information~\cite{ydjal19aamas,nt20ai,nt20aamas,nt21ijcai}. We also proposed a formal system for capturing trolley-like ethical dilemmas~\cite{ny21aaai}. For an effective responsibility ascription, delegation is a key concept in systems where multiple agents interact and act autonomously. Successfully delegating a task or goal incurs a transfer of responsibility, but the resulting mix of responsibilities flowing from this act in some social context can be complex. The party issuing the imperative is acting through another and, following the legal principle \textit{Qui facit per alium facit per se}, has acted themselves if the delegated task is achieved, and hence is responsible. Imperatives may be issued to groups of agents, leading to forms of collective responsibility. Grounded upon a formalisation of Hamblin's Action-State Semantics \cite{Hamblin1987,ReedNorman2007}, we have charaterised individual and group-directed imperatives in a rich notion of delegation \cite{NormanReed2010}.

\paragraph{Argumentation-Based Reasoning} Abstract argumentation frameworks provide rigorous formalisations to capture uncertainty about the true state of some situation. In their simple form, referred to as Dung Argumentation Frameworks (DAFs) \cite{Dung1995}, they consist of a set of abstract arguments and a binary defeats relation. These models capture qualitative uncertainties, and we can specify a variety of semantics to allow us to interpret a DAF to identify sets of arguments that are consistent to some standard. The use of these reasoning methods in the context of intelligence analysis was first demonstrated by Toniolo et al. \cite{Toniolo2015,Cerutti2018}, where structured argument maps developed by analysts could be mapped to a DAF and alternative interpretations (extensions) of the evidence (using preferred semantics) highlighted. Evaluation of the use of CISpaces with professional analysts provided good evidence that this promoted high-quality intelligence products, but we cannot, for example, capture uncertainties such as the trustworthiness of a source \cite{Parsons2011}, or the belief that some argument is valid. Probabilistic argumentation frameworks offer a solution by associating probabilities with abstract arguments \cite{Li2011} and, in combination with evidential frameworks \cite{Oren2008,Li2013}, they enable the translation of probabilistic uncertainties associated with evidence to abstract frameworks for reasoning. Experiments with humans have indicated that probabilistic approaches \cite{Li2011,Li2013} and bipolar models \cite{Oren2008,Li2013} naturally model some aspects of how we reason under uncertainty \cite{Polberg2018}.

\subsection{Open Challenges and Ways Froward}

Most current research on responsibility has focused on simple single-shot interactions. Multi-step interactions allow an agent to delay action until others have acted and then to act accordingly. In some situations, this might allow the agent to achieve the desired outcome without acting themselves or even revealing their intentions. We think that properly defining and studying responsibility in multi-step interactions is an interesting direction of the future research in formal reasoning about responsibility. 

Significant effort by AIC researchers~\cite{ydjal19aamas,nt20ai,nt20aamas,nt21ijcai} and others~\cite{bht06jelia,bht07tark,b11jal,llm14jlc,hp17rsl} has directed towards understanding of how agent's knowledge affects responsibility. As it is common in formal epistemology, the formal models of knowledge assume that everything ``known'' must be necessarily true. At the same time, in real-world setting, agents often base their decisions and, as a result, become responsible for consequences of these decisions, based on facts that are likely, but not necessarily true. Such facts are usually called ``beliefs''. Beliefs are often based on agents' {\em trust} into data and commitments of other agents. We think that studying how such trust and beliefs affect responsibility is another promising direction of research in formal studies of responsibility.

\section{Trusted and Trustworthy MAS}\label{sec:4}
Designing autonomous systems that are trustworthy and can be trusted is a key challenge. Deployment of such systems in the real world demands that they operate safely and in a coordinated manner and that they are able to recognise and compensate for faults and failures and work in partnership with humans. The AIC group is leading the UK's effort in building trustworthy autonomous systems (TAS). Over the last 20 years, the AIC group has been host to a range of projects that have developed some of the foundational tools for TAS. This includes the ALADDIN Project (2005-2011)\footnote{\url{https://www.ecs.soton.ac.uk/research/projects/357}}, the ORCHID Programme (2011-2016)\footnote{\url{http://www.orchid.ac.uk}}, and more recently the UKRI Trustworthy Autonomous Systems Hub (2020-present)\footnote{\url{http://www.tas.ac.uk}}. These are all highly multi-disciplinary projects, bringing together expertise from MAS, human-computer interaction, machine learning, mathematics and the social sciences, working closely with industrial partners from a range of sectors, including disaster response, defence and security, energy systems and transportation. Over the course of these projects, it became clear that multiagent systems could not be designed to simply focus on efficiency. Instead, the users of such systems were more concerned about their reliability and the ability to integrate them into a human-based process in a seamless way. This led us to focus on notions of trust and trustworthiness which is now the core agenda of the UKRI TAS Hub. 

In what follows, we elaborate on some of the research areas we have developed fundamental tools and techniques for, and outline some future challenges.

\subsection{AIC's Research Agenda on Trustworthy MAS}
We have been designing trustworthy and trusted multiagent systems for a number of application areas. Our research has been supported by multiple governmental and industrial organisations including BAE Systems, Thales, Dstl, Northrup Grumman, Secure Meters, QinetiQ and Boeing.

\paragraph{Energy Efficiency} Part of our previous effort was on applying AI for saving energy in smart homes~\cite{vytelingum2010trading,vytelingum2011agent,ramchurn2011agent}. We have developed algorithms to coordinate smart homes to flatten the load on the grid and to optimise storage of energy to maximise returns for users while minimising peaks on the grid. We have also designed fair algorithms for demand response to ensure energy-poor users are not priced out of markets. 

\paragraph{Trust Assessment and Trustworthy Swarms} We designed a simulation platform for human-swarm interaction experimentation and evaluated the performance of our proposed multiagent coordination algorithms for DCOPs, MCTS and coalition formation. We also designed algorithms to coordinate human-machine teams, teams of robots (UAVs) and emergency responders. Co-creation of use cases for human-swarm interaction with industry partners (Dstl \& Thales) has driven a user-centred interaction interface design for scalable swarms ~\cite{divband2021designing}, and engagement with experienced UAV operators has informed the requirements for trustworthy swarm systems~\cite{parnell2022trustworthy}. We also proposed methods for learning under uncertainty to assess the trustworthiness of autonomous agents given observable behaviour, and making trust-informed decisions \cite{BurnettSycaraNorman2010,BurnettSycaraNorman2011,Ceruttietal2015,SensoyYilmazNorman2016}.

\paragraph{Self-adaptive and Robust Swarms} Our research in swarm robotics has pushed the boundaries of fault-detection and adaptation with simple robots. We developed robot controllers for swarms to robustly detect a variety of \textit{a priori} unknown sensory-motor faults using a dynamic signature of normality to discriminate between normal behaviours and faults \cite{tarapore2015err,tarapore2017generic,tarapore2019fault}. We have also proposed rapid adaptation of robots to quickly recover from a variety of damages~\cite{cully2015robots}. We generated diverse repertoires of compensatory behaviours for the swarm following damages or changes in their environment \cite{bossens2020qed}. This is complemented by collaborative learning across the swarm, allowing the members of the swarm to share their experiences for faster adaptation \cite{bossens2021rapidly}.
    
\paragraph{Safe MAS} A key element of trustworthy multiagent systems is that they must operate within, or at least be cognizant of safety constraints. We have developed methods for safety-aware multiagent systems such that teams of agents minimise the severity of failures while sustaining a mission~\cite{GaspariniKollingbaumNorman2018}. This Normative Decentralised POMDP model uses a domain model that captures safety constraints for reward shaping and in the process of searching for safe policies. Human guidance can also play a key role in the process of learning a safe policy. We have explored active learning methods in safety-critical environments where agents balance safe exploration with costs associated with seeking human input \cite{KazantzidisAAMAS2022}. Human engagement in learning can help mitigate challenges of scarce environmental feedback, but learning in continuous environments brings additional challenges. Our research on Active Adaptive Perception \cite{Bossens2019} helps to overcome the inductive bias in learning spatio-temporal patterns through an architecture that learns when and how to modify and use a perception module. It is shown that emergent strategies are developed for when the memory is accurate. Another approach is to use multi-representational policies to share learning on policies \cite{Bossens2021a}. This approach aims to provide lifetime-scalable methods. It develops task capacity, a measure to determine the maximum number of tasks a policy can accurately solve. In addition, we have proposed a model-based reinforcement learning algorithm called Explicit Explore, Exploit, or Escape (E4), which extends the Explicit Explore or Exploit (E3) algorithm by allowing targeted policies for policy improvement across known states, discovery of unknown states, as well as safe return to known states \cite{Bossens2021b}. Theoretical results
show that E4 finds a near-optimal constraint-satisfying policy in polynomial time whilst satisfying safety constraints throughout the learning process. We also studied adversarial learning settings, where the data providing agent has an incentive to mislead the learner into a particular direction \cite{bishop2020optimal}. Finally, in AIC we study a number of applications within learning in MAS. In particular, in \cite{stein2020strategyproof}, reinforcement learning is used to allocate resources.

\subsection{Open Challenges and Ways Forward}

Despite our contributions, we are facing many issues in designing trustworthy autonomous systems including the preservation of human values and ethics, ensuring that autonomous systems are resilient and can be trusted to behave according to their designed objectives. Other remaining challenges are distributed optimisation under uncertainty and human-machine teaming where humans and machines can equally take control of tasks and direct each other. Another issue is establishing a trustworthy human-AI partnership where decisions made by both humans and machines in collaborative settings are fully tracked and trusted and the organisation of human and machine teams can be achieved with maximum effect. The issue of interpretable machine learning tools to help practitioners understand the outputs of deep learning systems also remains open and challenging. 

Furthermore, we are planning to design an integrated fault-detection and adaptation system. We are also keen to further develop the theory underlying sparse swarms \cite{tarapore2020sparse}, and demonstrate this in practice across terrestrial (forest), marine and aerial environments. Swarm robot hardware and coordination algorithms typically assume robots operating in very close proximity (on the scale of centimetres or meters apart). Such densities are not practical in many applications, in terms of deployment, monitoring and post-mission recovery. Robots of the swarm may need to operate autonomously several kilometres apart. 

We can achieve resilient and scalable augmentation of human perception and actuation using robot swarms but there are still many challenges. One of the main challenges is the explainablity of swarms. We need a way to understand the complex state of the swarms and generate useful explanations to ensure that the human operator is neither uninformed nor overwhelmed. Another key element is predicting future state of the system and ensuring that it shifts to high performance states. Once the current and the future states of the system are known, we need self-organised and adaptive control structures for agents to collectively configure themselves to achieve their goals. A further challenge is situation-based trust calibration for human-swarm systems to ensure the right level of trust that guarantees team performance. An additional aim is to make multi-robot systems more resilient; i.e. to tackle \emph{a priori} unknown challenges. These could be adapting to faults sustained by individual robots during operation, or adapting to changes in the operating environment. We need to start to explore the suitability and preparedness of MAS for complex applications such as disaster response, defence and security, energy systems, and healthcare.

\section{Human-Centred MAS}\label{sec:5}

Arguably, ``\textit{the AI systems' objective is to achieve what humans want}''~\cite{russell2019human} and, to that end, we need to  focus on the design and development of autonomous agents and multiagent systems that consider the social values and preferences of humans. The human-centred view in multiagent systems research builds on the sociotechnical perspective that, in most application domains, artificial agents need to work in collaboration with humans as a collective~\cite{DBLP:journals/cacm/JenningsMNRRRR14}. To achieve this, the aim is to form partnerships that are technically effective and trusted by the heterogeneous society of stakeholders~\cite{ramchurn2021trustworthy}.

\subsection{AIC's Research Agenda on Human-Centred MAS} \label{sec51}

Our research on human-centred multiagent systems expands around various application domains. In this section, we mainly elaborate on research on human-centred smart grid, user-centric mobility and transportation systems, and agent-based methods for participatory sensing. Moreover, we discuss our research focused on privacy of humans in the emerging Internet of Vehicles (IoV) and highlight  methods for consent negotiation.

\paragraph{Human-Centred Smart Grid} In~\cite{ramchurn2012putting}, we highlighted that safeguarding the quality of life of future generations is highly dependent on an efficient electricity grid and identified technical challenges that the fields of multiagent systems and decentralised AI can contribute to. As became evident in recent years, the smart grid is an electricity grid where the bidirectional flow of both electricity and information allows demand to be actively managed in real time, such that electricity can be generated at scale from intermittent renewable sources~\cite{vytelingum2011theoretical}. 
This vision was followed by practical tools for micro grid  management~\cite{vytelingum2010agent} and resulted in nationally implemented techniques that allowed users to  monitor and control their energy consumption, enabling user-centred energy management~\cite{ramchurn2011agent,auffenberg2015,auffenberg2017comfort}. 
This line of work also focused on how users and household-level energy prosumers (electricity 
consumer units with the capacity to produce) can interact with the grid and take part in energy trading in an automated manner~\cite{vytelingum2011agent,vytelingum2010trading}. In addition to national-level impact, our smart grid research~\cite{aic_smar_grid} identified open research challenges that will be discussed in Section~\ref{sec52}.

\paragraph{Urban Mobility and Human-Centred Internet of Vehicles} 
Our line of work on urban mobility and the human-centred Internet of Vehicles (IoV) relies on the idea that a ``key element of the IoV, is the citizen that should be central to the system and the prime motivator for its development''~\footnote{\url{http://www.orchid.ac.uk/smart-grid-2/}}. To that end, it is crucial to capture the preferences of mobility agents (being riders, drivers, or service providers), elicit them, and take them into account using realistic multiagent preference aggregation and incentive engineering methods.\footnote{See the UK's Urban Strategy for Future Mobility~\cite{DoTReport} and the recent report on the Future of Connected and Automated Mobility in the UK~\cite{ramchurn2021future}.}  In this context, we are active in developing multiagent techniques for effective management of intersections~\cite{worrawichaipat2021resilient,yazdanpanah2021formal,DBLP:conf/ijcai/Iwase0G22,behrad2022} as well as mechanisms that support mobility-enabling services (e.g.\ for an optimal distribution of charging stations for electric vehicles)~\cite{gerding2011online,robu2013online,zavvos2021comprehensive}. 
Finally, to support the transition towards on-demand mobility and shared mobility services, we developed adaptive pricing mechanisms~\cite{drwal2017adaptive} to incentivise the relocation of shared vehicles and introduced multiagent incentive engineering methods for ridesharing (together with partners from Toyota Motor Europe)~\cite{DBLP:conf/ijcai/Iwase0G21}.  Within the shared mobility domain, we also looked at how to balance different objectives (including social and environmental)~\cite{miya2022}, how to involve riders in the routing process~\cite{ong2022}, and how to account for the cost of walking in ridesharing~\cite{lucia2022}.

\paragraph{Multiagent Systems  for Social Good} The endeavour to mitigate climate change and adapt to its inevitable consequences requires a multidisciplinary approach~\cite{DBLP:conf/atal/YazdanpanahMJSS20}. To that end, multiagent techniques have the potential to address problems around coordinating pollution sensors and to facilitate the operation of large-scale Internet-of-Things (IoT) systems in a sustainable manner. Furthering our research agenda on \emph{participatory sensing}~\cite{balsamo2017wearable}, we developed multiagent methods for monitoring air pollution using low-cost mobile devices. Comparing against city-scale scenarios that use well-funded facilities, our participatory sensing method resulted in gathering 33.4\% more information~\cite{zenonos2015coordinating,zenonos2016algorithm,zenonos2018coordinating}. With regard to IoT systems, a critical issue is around the  operation of large-scale multi-sensor systems in a sustainable way and with minimal consumption of energy. In  recent work, conducted during the  first lockdown of the COVID-19 pandemic in the UK, we deployed a multiagent architecture for distributed sensing and showed that using mobile sensors is not only as effective as exploiting stationary units but also helps reducing device numbers, in turn, leading to a more sustainable IoT~\cite{cetinkaya2021distributed}.

\paragraph{Privacy and Meaningful Consent} A key challenge in developing human-centred multiagent systems is to maintain the privacy/efficiency trade-off. For instance, in large-scale vehicular systems~\cite{zavvos2021privacy}, it is crucial to ensure the privacy of users and avoid extracting what is considered to be sensitive data (e.g., riders' personal data or commuting routines). In this context, our research on \textit{autonomous negotiation} has been applied to address challenges around when, and in what contexts, autonomous agents will be be able to represent users~\cite{DBLP:conf/ijcai/BaarslagKGJG17} and on  managing permissions in mobile apps~\cite{DBLP:conf/atal/BaarslagAGAPGs17}. The permission management system developed includes agents that  autonomously negotiate potential agreements for the user, allows users to refine agreements, and learns users' preferences from such 
interactions. Moreover, we focused on \textit{consent negotiation} as a means for reaching  \textit{meaningful consent} in multiagent systems~\cite{schraefel2017internet,gomer2014consenting}. This approach complements the standard line of work on static \textit{preference aggregation} by allowing dynamic interaction with users and keeping humans in the loop.

\subsection{Open Challenges and Ways Forward}  \label{sec52}

While our past research has focused on representing the preferences and incentives of human users in a range of important application domains, it is still an open challenge to design large, decentralised AI systems that can be fully trusted by non-expert citizen end users. To address this challenge, we are working towards a vision of \emph{citizen-centric AI systems} (CCAIS)~\cite{ccaisWebsite}.

\begin{figure}[htb]
    \centering
    \includegraphics[width=0.75\textwidth]{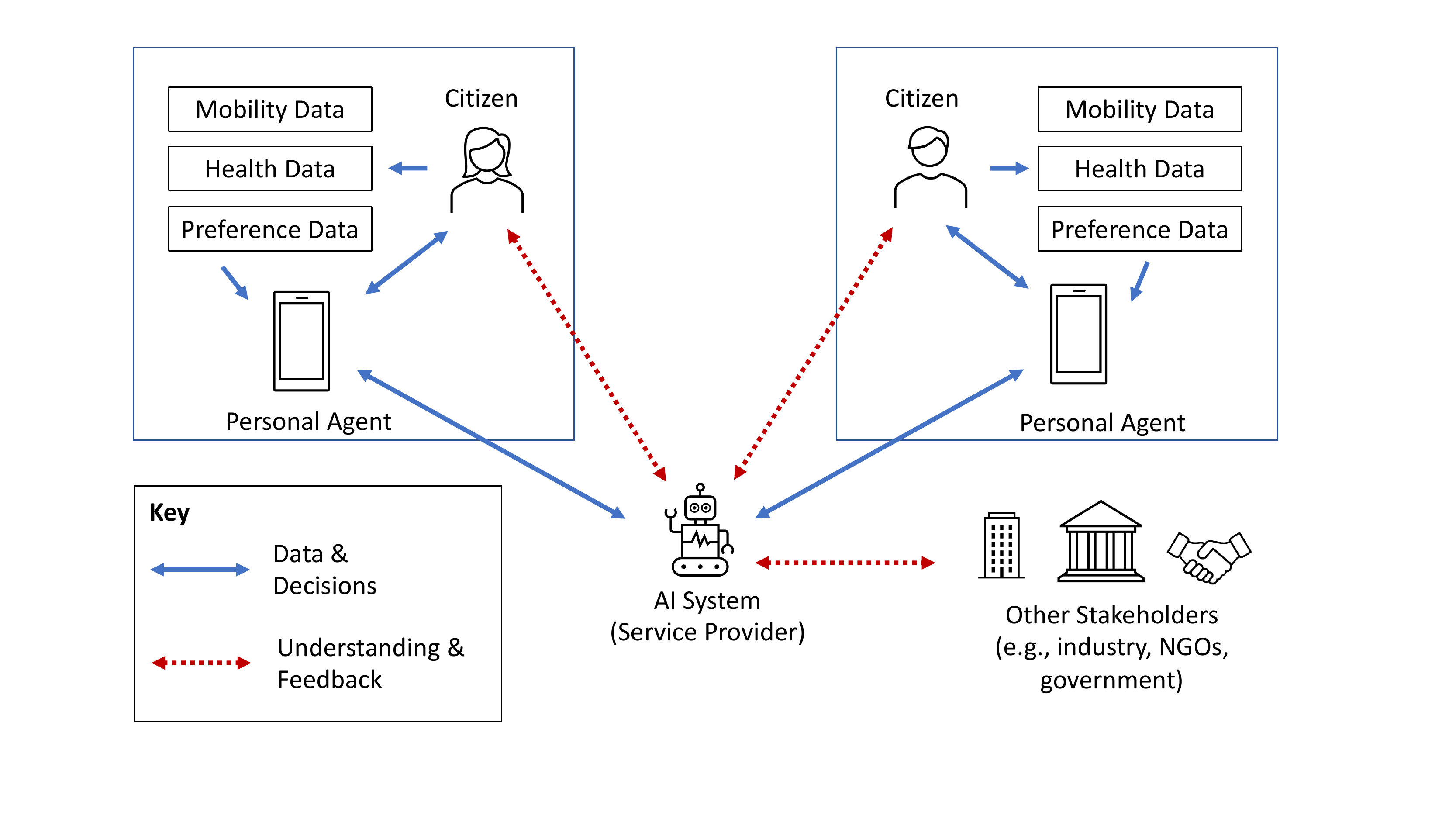}
    \caption{A Citizen-Centric AI System}
    \label{fig:ccais}
\end{figure}

As shown in Figure~\ref{fig:ccais}, these CCAIS are highly distributed multiagent systems, where personal intelligent agents represent the preferences and interests of individual citizens and negotiate with service providers on their behalf (e.g.,  a smart energy provider or a mobility-on-demand service). Importantly, these CCAIS meet a number of properties. They are: (1) \textit{citizen-aware}, i.e., through the use of personal agents, they understand the individual preferences and constraints of citizens; (2) \textit{citizen-beneficial}, i.e., through selective aggregation of preferences, negotiation and incentives, they are designed to benefit all citizens; (3) \textit{citizen-sensitive}, i.e.,  they are fair, inclusive and equitable by design; and (4) \textit{citizen-auditable}, i.e.,  they involve citizens and other stakeholders through a continuous feedback loop and offer clear explanations for their decisions. CCAIS research builds on the line of work around fairness-ensuring mechanisms and incentive engineering techniques (described in Section \ref{sec:2}), and aims at advancing the MAS research on both ends: (1) to keep humans in the loop, take a sociotechnical perspective, and look at fairness/incentives for humans while preserving technical efficiency; and (2) to focus on practical problems that human-AI collectives may face, but can solve collaboratively, and investigate challenges around implementing such a partnership. This sociotechnical perspective is not purely about  practical contributions and implementing available fairness and incentive theories but also introduces new theoretical challenges. 

\paragraph{Diverse and Inclusive MAS} 
In particular, we are interested in capturing the preferences and values of citizens in society in a diverse and inclusive form. This perspective contrasts with the historical view to develop services that are effective for a majority of \textit{generic} users, but which dismisses special cases.  To that end, a challenge is to elicit and model real-life conditions of citizens (e.g.\ physical diversities and cultural preferences) and investigate how such dynamics may  affect the efficacy of  incentivisation mechanisms and accordingly the reliability of a CCAIS.

\paragraph{Smart Energy Systems Under Deep Uncertainty} Building on research on smart grids and urban mobility (Section~\ref{sec51}), we are focused on addressing challenges around deeply uncertain energy markets and ensuring stability for energy neighbourhoods~\cite{buermann2020fair}. In neighbourhoods that aim for diversifying their energy sources, it is still an open challenge to decide what profitable coalitions to form and what energy profiles to implement. Considering more recent volatility in the energy market and calls for developing tools to address the challenge \cite{aiStrategy,smartGrid}, we are focused on developing decision support tools for energy cooperatives at the neighbourhood level and supporting citizens in ensuring their energy resilience.

\section{Conclusion}\label{sec:6}

The Agents, Interaction and Complexity group at the University of Southampton has a thriving community of academics, research staff and PhD students with a strong network of collaborators and alumni across the globe. The broad spectrum of research we have covered here includes theories of how self-interested agents interact and work toward fair outcomes, how agents reason and learn under uncertainty and how we can model and reason about MAS, developing trustworthy, safe and resilient autonomous and multiagent systems, and how, most importantly, humans are engaged at the centre of both our MAS research and the systems we conduct research into. The future of our multiagent systems research is an interdisciplinary endeavour in studying complex, long-term, evolving, and  trustworthy systems that are human-centred. In this way we believe our research has the potential to address global challenges such as sustainable development goals and progress societal benefits from AI.

\paragraph{Data Access Statement} 
Data sharing is not applicable to this article because no new data were created or analysed in this work. 

\paragraph{Acknowledgements} We acknowledge contributions of past members of the AIC group and, in particular, Professor Nick Jennings for leading AIC to become a world-leading research group in multiagent systems research. This work is supported by the UK Engineering and Physical Sciences Research Council (EPSRC) through 
the Trustworthy Autonomous Systems Hub (EP/V00784X/1), 
a Turing AI Acceleration Fellowship on Citizen-Centric AI Systems  (EP/V022067/1),
the platform grant entitled ``AutoTrust: Designing a Human-Centred Trusted, Secure, Intelligent and Usable Internet of Vehicles'' (EP/R029563/1), and by the UK Research and Innovation (UKRI) Centre for Doctoral Training in Machine Intelligence for Nanoelectronic Devices and Systems (EP/S024298/1). 
For the purpose of open access, the authors have applied a creative commons attribution (CC BY) licence to any accepted manuscript version arising.

\bibliographystyle{ios1}         
\bibliography{bibliography}      

\end{document}